\documentclass[
]{ceurart}

\usepackage{listings}
\usepackage{float}
\usepackage{graphicx}
\usepackage{todonotes}
\usepackage{amsmath}
\usepackage{multirow}
\usepackage{placeins}
\usepackage{graphicx}
\usepackage{tabularx}
\usepackage{pgfplots}
\usepackage{pgfplotstable}
\usepackage{tikz}
\usepackage{subcaption}   % for 'subfigure' environments
\pgfplotsset{compat=1.17}
\usepackage{pgfplots}
\usepackage{diagbox} % for diagonal splitting
\usepackage{float}   % for [H] specifier (optional)
\usepackage{changepage}

\begin{document}

%%
%% Rights management information.
%% CC-BY is default license.
\copyrightyear{2025}
\copyrightclause{Copyright for this paper by its authors.
  Use permitted under Creative Commons License Attribution 4.0
  International (CC BY 4.0).}

% DO NOT CHANGE THE TEMPLATE BELOW, but only the text
%%
%% This command is for the conference information
\conference{Late-breaking work, Demos and Doctoral Consortium, colocated with the 3rd World Conference on eXplainable Artificial
Intelligence: July 09–11, 2025, Istanbul, Turkey}

%%
%% The "title" command
\title{On the interplay of Explainability, Privacy and Predictive Performance with Explanation-assisted Model Extraction}

% \tnotemark[1]

% %%
% %% The "author" command and its associated commands are used to define
% %% the authors and their affiliations.
% \author[1,2]{Name1 Surname1}[%
% orcid=0000-0002-0877-7063,
% email=kulyabov-ds@rudn.ru,
% url=https://yamadharma.github.io/,
% ]
% \cormark[1]
% \fnmark[1]
% \address[1]{Peoples' Friendship University of Russia (RUDN University),
%   6 Miklukho-Maklaya St, Moscow, 117198, Russian Federation}
% \address[2]{Joint Institute for Nuclear Research,
%   6 Joliot-Curie, Dubna, Moscow region, 141980, Russian Federation}

% \author[3]{Name1 Surname1}[%
% orcid=0000-0001-7116-9338,
% email=i.tiddi@vu.nl,
% url=https://kmitd.github.io/ilaria/,
% ]
% \fnmark[1]
% \address[3]{Vrije Universiteit Amsterdam, De Boelelaan 1105, 1081 HV Amsterdam, The Netherlands}

% \author[4]{Name1 Surname1}[%
% orcid=0000-0002-9421-8566,
% email=Manfred.Jeusfeld@acm.org,
% url=http://conceptbase.sourceforge.net/mjf/,
% ]
% \fnmark[1]
% \address[4]{University of Skövde, Högskolevägen 1, 541 28 Skövde, Sweden}

% %% Footnotes
% \cortext[1]{Corresponding author.}
% \fntext[1]{These authors contributed equally.}
\author[1,2]{Fatima Ezzeddine}[%
email=fatima.ezzeddine@usi.ch,
]
\cormark[1]
\fnmark[1]
\address[1]{Università della Svizzera italiana, Lugano, Switzerland}
\address[2]{University of Applied Sciences and Arts of Southern Switzerland, Lugano, Switzerland}

\author[3]{Rinad Akel}[%
]
\fnmark[1]
\address[3]{Lebanese University, Beirut, Lebanon}

\author[3]{Ihab Sbeity}[%
]
%\address[3]{Lebanese University, Lebanon}

\author[2]{Silvia Giordano}[%
]
%\address[3]{University of Applied Sciences and Arts of Southern Switzerland, Switzerland}

\author[1]{Marc Langheinrich}[%
]
%\address[1]{Università della Svizzera italiana, Switzerland}

\author[2]{Omran Ayoub}[%
]
%\address[3]{University of Applied Sciences and Arts of Southern Switzerland, Switzerland}

%% Footnotes
\cortext[1]{Corresponding author.}
\fntext[1]{These authors contributed equally.}
%%
%% The abstract is a short summary of the work to be presented in the
%% article.
\begin{abstract}
Machine Learning as a Service (MLaaS) has gained important attraction as a means for deploying powerful predictive models, offering ease of use that enables organizations to leverage advanced analytics without substantial investments in specialized infrastructure or expertise. However, MLaaS platforms must be safeguarded against security and privacy attacks, such as model extraction (MEA) attacks. The increasing integration of explainable AI (XAI) within MLaaS has introduced an additional privacy challenge, as attackers can exploit model explanations—particularly counterfactual explanations (CFs) to facilitate MEA. In this paper, we investigate the trade-offs among model performance, privacy, and explainability when employing Differential Privacy (DP), a promising technique for mitigating CF-facilitated MEA. We evaluate two distinct DP strategies: implemented during the classification model training and at the explainer during CF generation.
\end{abstract}

\begin{keywords}
  Counterfactual Explanations \sep
  Model Extraction Attack \sep
  Differential Privacy
\end{keywords}

\maketitle

\section{Introduction}
Machine Learning (ML) as a Service (MLaaS) is becoming increasingly popular for deploying powerful predictive models as it facilitates access to ML training and deployment tools, while eliminating the need for extensive computational resources \cite{tramer2016stealing}. The adoption of MLaaS, however, introduces important security and privacy risks. For instance, adversaries can query the deployed ML models through application programming interfaces (APIs) to perform various types of attacks, such as membership inference (MIA) \cite{shokri2017membership} and model extraction (MEA) \cite{tramer2016stealing}. These attacks, if successful, pose serious threats to data privacy and intellectual property. For instance, MIA can reveal whether specific data points were used in training, MEA, instead, enables adversaries to replicate proprietary models, leading to financial losses and competitive disadvantages and facilitates other data privacy attacks by having access to a copy of the model.
To defend against these attacks, data privacy-enhancing technologies such as Differential Privacy (DP) \cite{dwork2006differential} exist. DP has shown effectiveness in defending against such attacks and is therefore widely adopted in use cases that require data and model sharing and deployments \cite{abadi2016deep}. DP enables privacy-preserving training of deep neural networks (DNN) to effectively mitigate inferential attacks by adding a controlled amount of noise to either raw data or model weights and ensures that individual data points have minimal influence on the model’s response, which limits the amount of sensitive information leaked when an attacker queries the model.

Recently, with the increasing demand for transparency in automated decision-making, MLaaS platforms are starting to incorporate Explainable Artificial Intelligence (XAI) \cite{guidotti2018survey} techniques into their workflows to provide explanations of the model's decisions \footnote{
https://aws.amazon.com/sagemaker/clarify/, 
https://cloud.google.com/explainable-ai} \cite{ezzeddine2024privacy,shokri2021privacy}. These platforms now provide not only the final decisions of ML models but also explanations of the underlying processes. The increased transparency provided by XAI introduces new challenges for preserving privacy and safeguarding MLaaS platforms from adversarial threats, as model's explanations can inadvertently reveal information about model's decision boundaries \cite{spartalis2023balancing}. Specifically, counterfactual explanations (CFs) \cite{wachter2017counterfactual}, which aim to identify the smallest changes to input data that would alter an ML model's prediction to a desired outcome, can reveal the factors most influential in the model's decision-making. Recent research has indeed explored how explanations can be leveraged to enhance the effectiveness of such attacks \cite{spartalis2023balancing,shokri2021privacy,ezzeddine2024knowledge,ezzeddine2024privacy,aivodji2020model}.
Complementing this, DP can also be applied at the explanation level, where it masks explanations to limit their utility to adversaries while balancing interpretability and privacy \cite{shokri2021privacy,ezzeddine2024knowledge}.
As DP can have impact on predictive performance and explanation quality, and can be applied on both levels, there is a growing research to highlight the importance of DP in developing mitigation strategies that specifically address risks introduced by explanations, emphasizing the need to adapt, utilize, or extend existing defense methods to the exploitation of explainability. 
In this work, we focus on analyzing the mitigation framework that integrates DP at the model and at the explainer, and investigate the interplay between $i)$ \emph{model's accuracy}, as DP is expected to influence model's inference capability, $ii)$ \emph{privacy}, as employing DP provides resilience against attacks, and $iii)$ \emph{explainability}, as noise added to model or explainer may impact quality of explanations \cite{ezzeddine2024privacy,abbasi2024further}. We aim to quantify this interplay and extract insights on the choice of where to employ DP (at model or at explainer, or at both) and the degree of noise level to be employed to balance predictive performance, explainability and privacy.

To perform the attack, we employ a recently proposed MEA technique based on Knowledge Distillation (KD) due to its proven performance and practicality \cite{ezzeddine2024knowledge}. In terms of mitigation strategies, we employ DP at the ML model using Differential Private-SGD (DP-SGD) and at the explainer using a DP-based Generative Adversarial Network (GAN) \cite{ezzeddine2024knowledge} with varying noise levels. To this end, we investigate the following research questions (RQs): %\footnote{Code will be made available upon acceptance}.
\begin{itemize}
    \item \emph{RQ1: To what extent does applying DP at the model or at the explainer, or both, effectively mitigates MEA facilitated by CFs?}
    \item \emph{RQ2: How does noise level in DP influence the effectiveness of MEAs that leverage CFs?}
    \item \emph{RQ3: In what ways does the quality of CF explanations differ when DP is applied at the model compared to the explainer?}
\end{itemize}

\section{Related Work}\label{relatedWork} % Fatima

Several studies have explores leveraging XAI techniques and exploiting model explanations to perform privacy attacks. In \cite{oksuz2023autolycus}, authors explore the vulnerabilities of Local Interpretable Model-agnostic Explanations and show that an adversary can generate new data samples near the decision boundary and, consequently, perform MEA by crafting adaptive queries. In \cite{shokri2021privacy}, authors show that by leveraging gradient-based explanations, adversaries can enhance the effectiveness of MIA. In \cite{yan2022towards}, the authors propose a methodology that performs MEA by jointly minimizing classification and explanation loss, thereby improving its fidelity. Other works explore the use of CFs to enhance the effectiveness of MEA. For instance, \cite{aivodji2020model} introduces a methodology that relies on model predictions and CFs to train a substitute model. Similarly, \cite{wang2022dualcf} presents a novel strategy where CF pairs, including the CF of the CF, serve as training samples to MEA. More recently, \cite{ezzeddine2024knowledge} proposes a methodology based on KD techniques that exploit CFs to perform MEA effectively while minimizing the number of queries to an MLaaS system and to generate private CFs with DP. Moreover, \cite{dissanayake2024model} explores the theoretical foundation of MEA with CFs highlighting the risks associated with providing CF explanations.

Several approaches have been proposed to prevent adversaries from exploiting model explanations for privacy attacks. In \cite{an2024counterfactual}, the authors propose an approach that builds on the concept of providing CFs that are not derived from the entire feature space but instead are generated within a designated space. Some works developed methodologies to generate explanations while limiting the exposure of sensitive insights related to decision boundaries, training data, or model architectures. Authors in \cite{yang2022differentially} present an approach to generate differentially private CFs using functional mechanisms to protect the underlying model from potential inference attacks. In contrast, \cite{pentyala2023privacy} proposes a novel approach that constructs private recourse paths as CFs using differentially private clustering. Authors in \cite{ezzeddine2024knowledge} focus on GAN-based CF (proposed in \cite{nemirovsky2020countergan}), injecting DP into the training process of the generator that is responsible for generating CFs that limits the memorization of the private data points.

Similar to these works, we focus on identifying a mitigation strategy against attacks that exploit the model's explanations. Specifically, we explore the application of DP to the ML model, the explainer, and both simultaneously. Despite the numerous studies utilizing DP for mitigation strategies, our work is, to the best of our knowledge, the first to explore the application of DP in both the ML model and the explainer and to investigate their effectiveness in countering MEA and examine their influence on the quality of explanations. Additionally, our work explores the interplay between preserving model privacy and generating privacy-preserving CFs, as well as the implications for defending against MEA.

\section{Problem Formulation and Methodology}\label{Methodology} 
\vspace{-17pt}
\begin{figure}[ht]
\centering
    \includegraphics[scale=0.34]{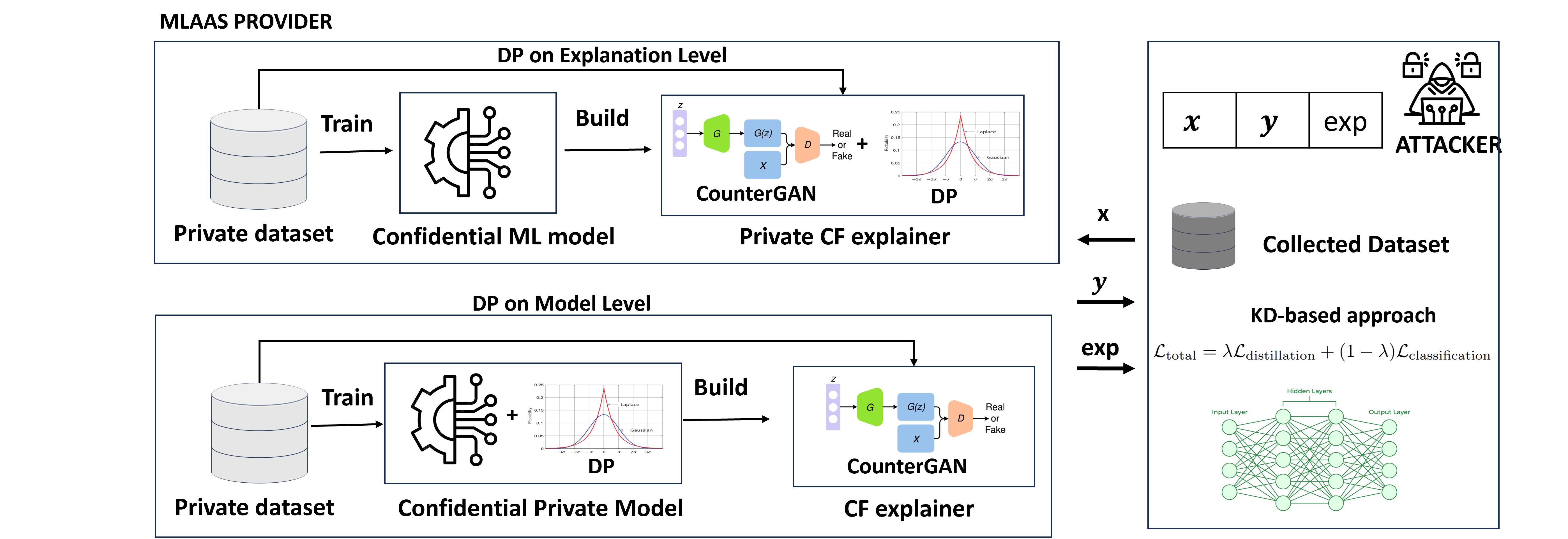}
    \caption{Model Extraction Attack within an MLaaS provider, depicting two different scenarios where DP is employed at the model or at the explainer to counter potential attacks.}
    \label{fig:overallframework}
\end{figure}
Given a dataset $D = \{(x_i, y_i)\}_{i=1}^{N}$, where $x_i$ are feature vectors and $y_i$ are corresponding labels. A target model $f(x; \theta)$ trained and optimized to achieve high performance on $D$ is deployed as MLaaS and is queryable through an API (as shown in Fig. \ref{fig:overallframework}). An ${Attacker}$ (adversary) attempts to extract an approximation of $f(x; \theta)$ using queries and the provided CFs. The attacker conducts MEA by exploiting CFs and varying the number of queries. To perform our analysis, we proceed as follows:
\begin{itemize}
    \item Step 1: Train target models as baseline models $f_{\text{baseline}}(x; \theta_{\text{base}})$.
    \item Step 2: Generate CFs by training a CounterGAN to generate CFs $\hat{x} = G(x; \phi)$ for $f_{\text{baseline}}$.
    \item Step 3: Simulate MEA, where the adversary queries the models with random points and collects pairs of predictions and CFs. The adversary trains an extracted model using the KD-based method proposed in \cite{ezzeddine2024knowledge}.
    \item Step 4: Measure MEA success by computing the agreement on a separate dataset to quantify and compare the level of agreement between extracted and original models/explanations.
    \item Step 5: Assess the quality of CFs using metrics such as prediction gain, realism (explained in more details in Sec \ref{ExperimentalSettings}).
\end{itemize}

The effectiveness of the MEA is measured using similarity metrics such as agreement. In practice, this agreement expectation is estimated empirically using a set of \( n \) test inputs \( \{x_1, x_2, \dots, x_n\} \). $Agreement = \frac{1}{n} \sum_{i=1}^{n} {I} \left( f_{\theta}(x_i) = \hat{f}_{\hat{\theta}}(x_i) \right)$. Where \( {I} \) counts the number of times the extracted model’s predictions match the target model’s predictions.

As a mitigation against MEA, we employ two strategies: \emph{1) DP-Model with DP-SGD:} where we apply DP-SGD during model training on $f(x; \theta)$. \emph{2) DP-Explainer (DP in CounterGAN):} We inject DP noise at the generator $G(x; \phi)$ that outputs private CFs (\cite{ezzeddine2024knowledge}). We then perform MEA leveraging CFs under different DP settings, i.e., the approach adopted and the privacy parameter's noise level $\epsilon$, and evaluate the adversary’s MEA success and CF quality.  Specifically in step 1, $f_{\text{baseline}}(x; \theta_{\text{base}})$ is first trained on $D$ without DP. We also train a DP-protected model $f_{\text{DP}}(x; \theta_{\text{DP}})$ using DP-SGD with privacy parameter $\epsilon$. Similarly, in step 2, we also train a private CounterGAN $\hat{x} = G_{private}(x; \phi)$ to generate the private CFs by varying the noise level. The attacker in step 3, apply MEA to extract $f_{\mathcal{A}}(x; \theta_{\mathcal{A}})$ using the KD-based method using either CFs generated by $G(x; \phi)$ or $G_{private}(x; \phi)$. For the comparative analysis, we consider four distinct scenarios: \emph{(1) No DP:} Baseline scenario that does not incorporate DP at any level, allowing the evaluation of the unprotected model performance and vulnerability.  \emph{(2) DP-Model:} Only the target model employs DP. This protects the model from adversarial replication while the explanation generator remains unprotected.  \emph{(3) DP-Explainer:} DP is applied to the explanation generator. This scenario assesses the impact of DP explanations on their utility without directly affecting the target model. \emph{(4) DP-Model-Explainer:} Both the target model and the explanation generator are protected with DP, aiming to balance model performance, explanation quality, and resistance to MEA. 

\section{Experimental Settings}\label{ExperimentalSettings} %Rinad

\subsection{Datasets, Target and Threat Model}
We perform an evaluation on 2 datasets: \emph{Housing} \cite{scikit-learn-california} and \emph{EEG Eye State}\cite{eeg_eye_state}. The Housing dataset describes housing prices and includes 20,640 instances and 8 features a mix of socio-economic, demographic, and geographic attributes. The target variable represents the median house value and is converted into two classes using a threshold defined by the median. The EEG Eye State dataset comprises EEG measurement data recorded using a Neuroheadset, and contains 14,980 data points and 14 features. The target variable is a binary label representing the eye-closed or opened state. 

The target model \( f_{\theta} \) is a DNN with 16 hidden layers of 64, 32, 16, 32, 64, 128, 64, 32, 128, 64, 128, 64, 128, 64, 32, and 16 neurons per layer with a GELU activation function and a the softmax activation function in the output layer. We employ Adam optmizer for the cases where DP is not used and TensorFlow Privacy’s DPKerasAdamOptimizer for the cases where DP applied. The model is trained without DP and with noise levels of 0.1, 0.5 and 0.9 for DP cases and with varying learning rates (0.001, 0.002, and 0.01), and we vary the l2\_norm\_clip to between 1, and 1.5 (l2\_norm\_clip bounds the sensitivity of the gradients by limiting the influence of any single training example on the overall gradient update, which is a crucial step before adding noise). Note that the more noise, the higher the privacy. The target models are trained using 80\% of the corresponding dataset, and the best-performing model in term of accuracy was chosen.

To simulate a realistic attack scenario, we assume that the attacker has no prior knowledge of the training data distribution and does not know the architecture of the target model, but can build a simple threat model \( t_{\Upsilon} \). The \( t_{\Upsilon} \) consists of 5 layers, with 32, 64, 128, and 64 neurons with ReLU activation, followed by a softmax output layer. Attacker generate random different data points to query the model, within a range of -3 to 3 for each feature and extract CFs to feed as input to the KD-based MEA.  Our evaluation involves performing MEA while varying the number of queries from 50 to 1000 and therefore the input to KD. For optimization, we utilize both Adam for the cases where DP is not used and TensorFlow Privacy’s DPKerasAdamOptimizer for the cases where DP is used to assess model performance under three different noise levels, 0.1, 0.5, and 0.9. We tune different KD-based approach hyperparameters, specifically, alpha within the range of 0.1 and 0.5, temperature within the range of 1 and 10. We compute the MEA agreement over the 20\% test set and report average results of 5 runs.

\subsection{Counterfactual generator}
The generator of CounterGAN takes an input feature vector and processes it through 4 layers with 64, 32, and 64 neurons, with ReLU as an activation function and a final layer with Tanh activation. The discriminator follows a simple feedforward design, consisting of 128, 128, 64 neurons with ReLU activation, and the final output layer with Sigmoid activation. In the No-DP scenario, we used the standard Adam optimizer. For the scenarios where DP is employed, we applied DP using noise levels of 0.1, 0.5, and 0.9, respectively on the generator, with TensorFlow Privacy’s DPKerasAdamOptimizer. We varied the learning rate where we used 0.05, 0.005, 0.01, and 0.001, and l2\_norm\_clip to between 1, 1.5, and 3. We report the results of the average of 5 runs.
We consider the following metrics to assess the influence of employing privacy on the CFs. 
%To evaluate the quality of the CFs, we compute the following metrics:
\begin{itemize}
    \item Prediction Gain: quantifies how the explainer modifies the input to influence the model's decision by measuring the change in the classifier’s confidence score for a specific target class \( t \) when replacing the original data point with its CFs: $\Delta P = P_{f}(CF, t) - P_{f}(X, t)$ Where: \( P_{f}(CF, t) \) is the probability score for the target class \( t \) of the CF and \( P_{f}(X, t) \) is for initial point. %By analyzing prediction gain, we can determine whether the CFs successfully shift the model decisions toward the desired target class.

    \item Realism: quantifies how a data instance fits within a data distribution to evaluate how well CFs and private CFs with different noise applied match the original training data distribution. It is defined as: $\text{Realism} = \frac{1}{N} \sum_{i=1}^{N} \| \text{input}_i - \text{reconstruction}_i \|^2$
    Where: \( \text{input}_i \) represents the original data point, \( \text{reconstruction}_i \) is the corresponding autoencoder reconstruction and \( N \) is the total number of instances (\cite{nemirovsky2020countergan}). A lower realism value indicates that the data point is more realistic. 
\end{itemize}

\section{Results and Discussion}\label{Results}
\subsection{ML Model Predictive Performance}\label{ModelPerf} 
\vspace{-10pt}
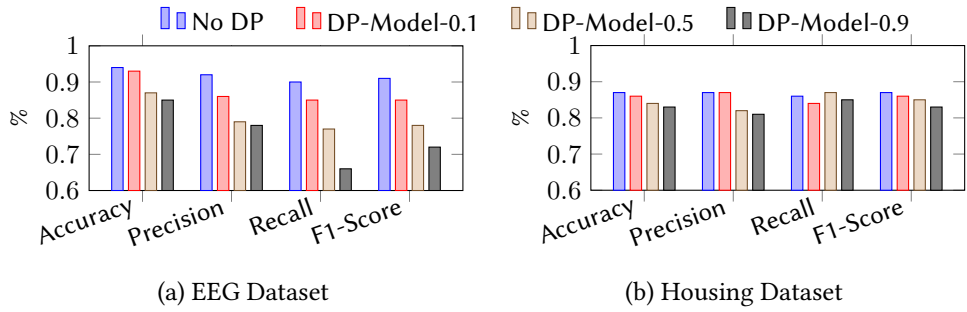
\begin{figure}[ht]
  \centering
    \begin{minipage}{0.4\textwidth} % First plot
        \centering
        \begin{tikzpicture}
            \begin{axis}[
                ybar,
                bar width=0.15cm,
                width=6.5cm,
                height=3.5cm,
                symbolic x coords={Accuracy, Precision, Recall, F1-Score},
                xticklabel style={font=\small, rotate=20, anchor=east},
                xtick=data,
                ymin=0.6, ymax=1.0,
                ylabel={\%},
                legend style={at={(1.2,1)}, anchor=south, legend columns=-1, draw=none, 
              /tikz/every even column/.append style={column sep=10pt}},
                enlarge x limits=0.2,
            ]
                \addplot coordinates {(Accuracy,0.94) (Precision,0.92) (Recall,0.90) (F1-Score,0.91)};
                \addplot coordinates {(Accuracy,0.93) (Precision,0.86) (Recall,0.85) (F1-Score,0.85)};
                \addplot coordinates {(Accuracy,0.87) (Precision,0.79) (Recall,0.77) (F1-Score,0.78)};
                \addplot coordinates {(Accuracy,0.85) (Precision,0.78) (Recall,0.66) (F1-Score,0.72)};
                \legend{No DP\hspace{40pt}, DP-Model-0.1\hspace{10pt}, DP-Model-0.5\hspace{10pt}, DP-Model-0.9}
            \end{axis}
        \end{tikzpicture}
         \subcaption{EEG Dataset}
    \end{minipage}
    \begin{minipage}{0.4\textwidth} % Second plot
        \centering
        \vspace{10pt}
        \begin{tikzpicture}
            \begin{axis}[
               ybar,
                bar width=0.15cm,
                width=6.5cm,
                height=3.5cm,
                symbolic x coords={Accuracy, Precision, Recall, F1-Score},
                xticklabel style={font=\small, rotate=20, anchor=east},
                xtick=data,
                ymin=0.6, ymax=1.0,
                ylabel={\%},
                enlarge x limits=0.2,
            ]
                \addplot coordinates {(Accuracy,0.87) (Precision,0.87) (Recall,0.86) (F1-Score,0.87)};
                \addplot coordinates {(Accuracy,0.86) (Precision,0.87) (Recall,0.84) (F1-Score,0.86)};
                \addplot coordinates {(Accuracy,0.84) (Precision,0.82) (Recall,0.87) (F1-Score,0.85)};
                \addplot coordinates {(Accuracy,0.83) (Precision,0.81) (Recall,0.85) (F1-Score,0.83)};
            \end{axis}
        \end{tikzpicture}
        \subcaption{Housing Dataset}
    \end{minipage}
    \caption{Model Performance achieved by the ML model across the two datasets for varying noise scales.}
    \label{tab:ModelPerformance}
\end{figure}
\vspace{-15pt} 
Fig \ref{tab:ModelPerformance} reports the predictive performance metrics of the models while varying the noise level across the two datasets used in our evaluations. As previously mentioned, we consider three noise levels when applying DP, 0.1, 0.5 and 0.9, and we refer to each case as DP-Model-{noise level}. As expected, the results across the two datasets indicate a decline in predictive performance metrics as the noise level increases. For instance, in the EEG dataset, accuracy, precision, recall, and F1-score are 0.94, 0.92, 0.9, and 0.91, respectively, when no DP is applied. However, at the highest noise level considered (0.9), these metrics drop to 0.85, 0.78, 0.66, and 0.72, respectively. Similar results are seen across the Housing dataset, where predictive performance metrics show a declining trend as the noise level applied increases.

\subsection{Effectiveness of Differential Privacy in Mitigating MEA}\label{AgreementRes}
 
\begin{figure}[htb]
  \centering
   \hspace*{-2.5em}%
  \begin{minipage}[t]{0.24\textwidth}
    \centering
    \begin{tikzpicture}[baseline]
      \begin{axis}[
        width=\linewidth,
        height=4cm,
        xlabel={Queries Number},
        ylabel={Agreement},
        legend columns=4,
        legend style={
          at={(3,1.1)},
          anchor=south,
          font=\small,
          draw=none,
          fill=none,
          inner sep=1pt
        },
        ymin=45, ymax=85,
        grid=major,
        xtick={50,100,200,300,500,1000},
        ytick={40,50,60,70,80,90},
        xticklabel style={rotate=90,font=\tiny},
      ]
        \addplot[color=red,mark=square*,thick] coordinates {
            (50, 70.32) (100, 71.17) (200, 74.87) (300, 76.4) (500, 79.23) (1000,80.23)
        };
        \addlegendentry{No DP}
        \addplot[color=blue,mark=o,thick] coordinates {
            (50,59.4) (100, 65.9) (200, 68.21) (300, 70.49) (500, 73.13) (1000,77.54)
        };
        \addlegendentry{DP-Explainer-0.1}
        \addplot[color=green,mark=triangle*,thick] coordinates {
            (50, 62.64) (100, 64.34) (200, 66.1) (300, 68.9) (500, 75.79) (1000,76.45)
        };
        \addlegendentry{DP-Explainer-0.5}
        \addplot[color=purple,mark=diamond*,thick] coordinates {
            (50, 49.22) (100, 50.7) (200, 58.43) (300, 59.8) (500, 64.6) (1000,74.41)
        };
        \addlegendentry{DP-Explainer-0.9}
      \end{axis}
    \end{tikzpicture}
    \subcaption{No DP}
  \end{minipage}\hspace{-0.5em}
  \begin{minipage}[t]{0.24\textwidth}
    \centering
    \begin{tikzpicture}[baseline]
      \begin{axis}[
        width=\linewidth,
        height=4cm,
        xlabel={Queries Number},
        ymin=50, ymax=85,
        grid=major,
        xtick={50,100,200,300,500,1000},
        ytick={40,50,60,70,80,90},
        xticklabel style={rotate=90,font=\tiny},
      ]
        \addplot[color=red,mark=square*,thick] coordinates {
            (50, 60.46) (100, 62.62) (200, 66.69) (300, 68.3) (500, 72.79) (1000,75.65)
        };
        \addplot[color=blue,mark=o,thick] coordinates {
            (50, 51.9) (100, 59.68) (200, 62.08) (300, 67) (500, 68.36) (1000,70.63)
        };
        \addplot[color=green,mark=triangle*,thick] coordinates {
            (50, 50.6) (100, 54.65) (200, 57.24) (300, 58.5) (500, 66.6) (1000,70.54)
        };
        \addplot[color=purple,mark=diamond*,thick] coordinates {
            (50, 54.45) (100, 54.48) (200, 60.24) (300, 64.65) (500, 65.6) (1000,70.27)
        };
      \end{axis}
    \end{tikzpicture}
    \subcaption{DP-Model-0.1}
  \end{minipage}\hspace{-0.5em}
  \begin{minipage}[t]{0.24\textwidth}
    \centering
    \begin{tikzpicture}[baseline]
      \begin{axis}[
        width=\linewidth,
        height=4cm,
        xlabel={Queries Number},
        ymin=55, ymax=75,
        grid=major,
        xtick={50,100,200,300,500,1000},
        ytick={40,50,60,70,80,90},
        xticklabel style={rotate=90,font=\tiny},
      ]
        \addplot[color=red,mark=square*,thick] coordinates {
            (50, 58.18) (100, 59.76) (200, 66.15) (300, 69) (500, 71.5) (1000, 73.71)
        };
        \addplot[color=blue,mark=o,thick] coordinates {
            (50,62.64) (100, 64.4) (200, 66.64) (300, 67.7) (500, 68.16) (1000,71.5)
        };
        \addplot[color=green,mark=triangle*,thick] coordinates {
            (50, 57.6) (100, 59.1) (200, 63.29) (300, 63.4) (500, 67.65) (1000,69.11)
        };
        \addplot[color=purple,mark=diamond*,thick] coordinates {
            (50, 58.6) (100, 61.38) (200, 62.28) (300, 65.35) (500, 67.68) (1000,68.89)
        };
      \end{axis}
    \end{tikzpicture}
    \subcaption{DP-Model-0.5}
  \end{minipage}\hspace{-0.5em}
  \begin{minipage}[t]{0.24\textwidth}
    \centering
    \begin{tikzpicture}[baseline]
      \begin{axis}[
        width=\linewidth,
        height=4cm,
        xlabel={Queries Number},
        ymin=50, ymax=75,
        xtick={50,100,200,300,500,1000},
        ytick={40,50,60,70,80,90},
        grid=major,
        xticklabel style={rotate=90,font=\tiny},
      ]
        \addplot[color=red,mark=square*,thick] coordinates {
            (50, 59.9) (100, 59.73) (200,65.69) (300, 67.61) (500, 70.1) (1000,71.7)
        };
        \addplot[color=blue,mark=o,thick] coordinates {
            (50, 56.85) (100, 61.81) (200,62.2) (300, 64.9) (500, 67.53) (1000,69.06)
        };
        \addplot[color=green,mark=triangle*,thick] coordinates {
            (50, 54.38) (100, 57.17) (200,59.95) (300, 61.45) (500, 62.83) (1000,67.12)
        };
        \addplot[color=purple,mark=diamond*,thick] coordinates {
            (50, 57.75) (100, 58.11) (200,60.99) (300, 62.88) (500, 63.66) (1000,64.29)
        };
      \end{axis}
    \end{tikzpicture}
    \subcaption{DP-Model-0.9}
  \end{minipage}
  \caption{Housing dataset MEA agreement for various DP strategies, noise levels and number queries.}
  \label{fig:CaliforniaAgreement}
\end{figure}
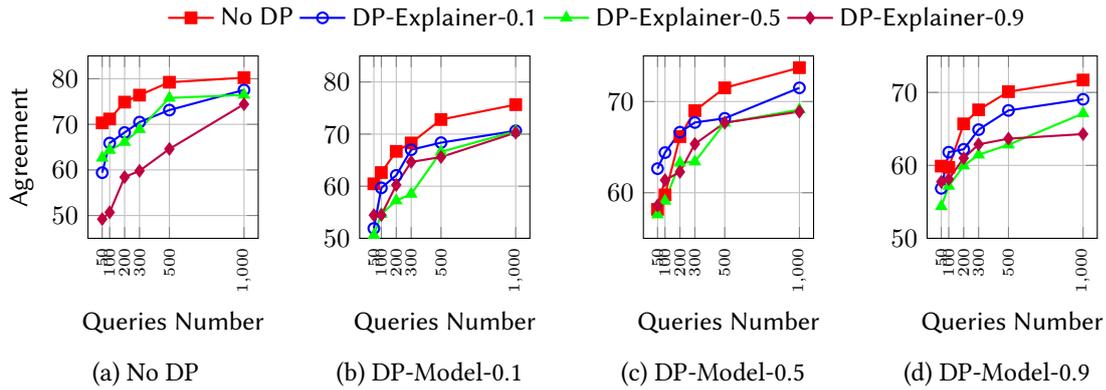
\begin{figure}[htb]
  \centering
  \hspace*{-2.5em}%
  \begin{minipage}[t]{0.24\textwidth}
    \centering
    \begin{tikzpicture}[baseline]
      \begin{axis}[
        width=\linewidth,
        height=4cm,
        xlabel={Queries Number},
        ylabel={Agreement},
        legend columns=4,
        legend style={
          at={(3,1.1)},
          anchor=south,
          font=\small,
          draw=none,
          fill=none,
          inner sep=1pt
        },
        ymin=40, ymax=100,
        grid=major,
        xtick={50,100,200,300,500,1000},
        ytick={40,50,60,70,80,90,100},
        xticklabel style={rotate=90,font=\tiny},
      ]
        \addplot[color=red,mark=square*,thick] coordinates {
            (50,68.68) (100, 70.4) (200,72.2) (300,87.41) (500,90.6) (1000,96.4)
        };
        \addlegendentry{No DP}
        \addplot[color=blue,mark=o,thick] coordinates {
            (50,65.32) (100, 69.34) (200,73.5) (300, 75.14) (500, 79.41) (1000,82.17)
        };
        \addlegendentry{DP-Explainer-0.1}
        \addplot[color=green,mark=triangle*,thick] coordinates {
            (50,63.4) (100, 67.9) (200,70.39) (300, 73.66) (500, 75.11) (1000,77.42)
        };
        \addlegendentry{DP-Explainer-0.5}
        \addplot[color=purple,mark=diamond*,thick] coordinates {
            (50,54.57) (100, 55.6) (200,61.18) (300, 66.8) (500, 69.31) (1000,74.06)
        };
        \addlegendentry{DP-Explainer-0.9}
      \end{axis}
    \end{tikzpicture}
    \subcaption{No DP}
  \end{minipage}\hspace{-0.5em}
  \begin{minipage}[t]{0.24\textwidth}
    \centering
    \begin{tikzpicture}[baseline]
      \begin{axis}[
        width=\linewidth,
        height=4cm,
        xlabel={Queries Number},
        ymin=40, ymax=85,
        grid=major,
        xtick={50,100,200,300,500,1000},
        ytick={40,50,60,70,80,90},
        xticklabel style={rotate=90,font=\tiny},
      ]
        \addplot[color=red,mark=square*,thick] coordinates {
            (50,57.21) (100, 60.32) (200,61.69) (300, 62.46) (500, 70.77) (1000,78.08)
        };
        \addplot[color=blue,mark=o,thick] coordinates {
            (50,51.6) (100, 52.07) (200,56.83) (300, 57.21) (500, 58.54) (1000,66.13)
        };
        \addplot[color=green,mark=triangle*,thick] coordinates {
            (50,52.39) (100, 53.79) (200,54.77) (300, 56.69) (500, 57.35) (1000,59.97)
        };
        \addplot[color=purple,mark=diamond*,thick] coordinates {
            (50,45.6) (100, 50.7) (200,51.17) (300, 51.66) (500, 53.68) (1000,57)
        };
      \end{axis}
    \end{tikzpicture}
    \subcaption{DP-Model-0.1}
  \end{minipage}\hspace{-0.5em}
  \begin{minipage}[t]{0.24\textwidth}
    \centering
    \begin{tikzpicture}[baseline]
      \begin{axis}[
        width=\linewidth,
        height=4cm,
        xlabel={Queries Number},
        ymin=50, ymax=75,
        grid=major,
        xtick={50,100,200,300,500,1000},
        ytick={40,50,60,70,80,90},
        xticklabel style={rotate=90,font=\tiny},
      ]
        \addplot[color=red,mark=square*,thick] coordinates {
            (50,58.44) (100, 59.97) (200,60.39) (300, 61.79) (500, 62.14) (1000,70.01)
        };
        \addplot[color=blue,mark=o,thick] coordinates {
            (50,54.8) (100, 56.03) (200,56.2) (300, 58.64) (500, 62.04) (1000,62.88)
        };
        \addplot[color=green,mark=triangle*,thick] coordinates {
            (50,51.36) (100, 54.46) (200,56.04) (300, 57.02) (500, 57.21) (1000,59.24)
        };
        \addplot[color=purple,mark=diamond*,thick] coordinates {
            (50,50.31) (100, 53.87) (200,56.3) (300, 56.51) (500, 59.24) (1000,59.94)
        };
      \end{axis}
    \end{tikzpicture}
    \subcaption{DP-Model-0.5}
  \end{minipage}\hspace{-0.5em}
  \begin{minipage}[t]{0.24\textwidth}
    \centering
    \begin{tikzpicture}[baseline]
      \begin{axis}[
        width=\linewidth,
        height=4cm,
        xlabel={Queries Number},
        ymin=40, ymax=75,
        grid=major,
        xtick={50,100,200,300,500,1000},
        ytick={40,50,60,70,80,90},
        xticklabel style={rotate=90,font=\tiny},
      ]
        \addplot[color=red,mark=square*,thick] coordinates {
            (50,55.99) (100, 56.16) (200,57.32) (300, 62.25) (500, 66.79) (1000,67.45)
        };
        \addplot[color=blue,mark=o,thick] coordinates {
            (50,50.92) (100, 53.26) (200,57.95) (300, 58.08) (500, 60.43) (1000,63.26)
        };
        \addplot[color=green,mark=triangle*,thick] coordinates {
            (50,49.96) (100, 50.27) (200,51.02) (300, 57.56) (500, 59.03) (1000,60.85)
        };
        \addplot[color=purple,mark=diamond*,thick] coordinates {
            (50,46.8) (100, 48.38) (200,51.2) (300, 52.18) (500, 56.76) (1000,58.65)
        };
      \end{axis}
    \end{tikzpicture}
    \subcaption{DP-Model-0.9}
  \end{minipage}
  \caption{EEG dataset MEA agreement for various DP strategies, noise levels and number queries.}
  \label{fig:EEGAgreement}
\end{figure}
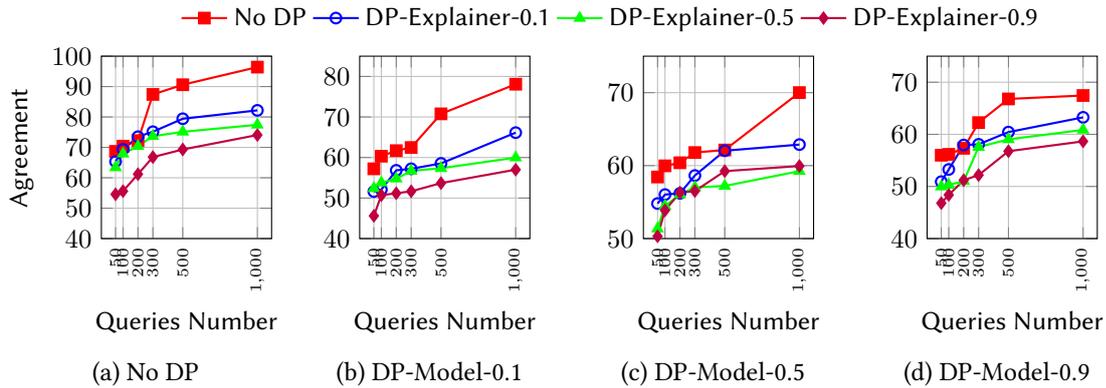
We considering the 3 scenarios of application of DP, namely, DP-Model, DP-Explainer and DP-Model-Explainer, and the baseline No DP scenario. Additionally, when we incorporate DP at explainer, we refer to each case as DP-Explainer-{noise level} (i.e., DP-Explainer-0.1). This evaluation will allow us to address RQ1 and RQ2. Figures \ref{fig:CaliforniaAgreement} show the \emph{agreement} observed by MEA across the various combinations of applying DP for varying noise levels and number of queries across the \emph{Housing} dataset. We start with No DP (Fig. \ref{fig:CaliforniaAgreement}(a)), which allows us to quantify solely the impact of employing different levels of noise at the explainer on the success of the MEA. Results show a general trend where the MEA is more successful as the number of queries used increases across all cases (i.e., independent of the noise level applied). Comparing the \emph{agreement} when employing different noise levels, results show that employing more noise, as expected, provides more defense against MEA. Specifically, with a noise level of 0.9, \emph{agreement} ranges between 50 and 72 when the number of queries increases up to 1000. In contrast, when employing noise levels of 0.5 and 0.1, \emph{agreement} falls within the ranges of 62–75 and 60–78, respectively. In the absence of DP at the explainer, \emph{agreement} starts at 70 with 50 queries and reaches 80 when 1000 queries are used. We now focus on the cases where DP is employed at the model level (Fig. \ref{fig:CaliforniaAgreement}(b), (c) and (d)). Generally, results show a similar trend across all cases, where \emph{agreement} increases with the number of queries used to perform the MEA. Comparing the \emph{agreement} achieved when employing different noise levels in each case, results show, as expected, that employing higher noise levels at the explainer implies better protection against MEA. For instance, when employing DP-Model with a noise level of 0.1 (Fig. \ref{fig:CaliforniaAgreement}(b)), the highest \emph{agreement} observed is 70 when DP-Explainer is employed (which is a DP-Model-Explainer case), compared to 76 without DP-Explainer. Similarly, with a noise level of 0.5 at the model (Fig. \ref{fig:CaliforniaAgreement}(c)), the agreement consistently remains lower than in the No DP case, reaching a maximum of 70.63 versus 75 to when DP is only applied at the model. Similar trends were observed to DP-Model-0.9.
Figure \ref{fig:EEGAgreement} show the agreement observed by MEA on the EEG dataset across the various cases. The results show similar trends to the one observed with the Housing Dataset. When no DP is applied to the model (Fig. \ref{fig:EEGAgreement}(a)), the \emph{agreement} improves with more queries ranging between 68\% and 96\%, with the highest agreement observed when DP is not applied at all. 

\subsection{Impact of Differential Privacy on Quality of Explanations}\label{CFAnalysis}
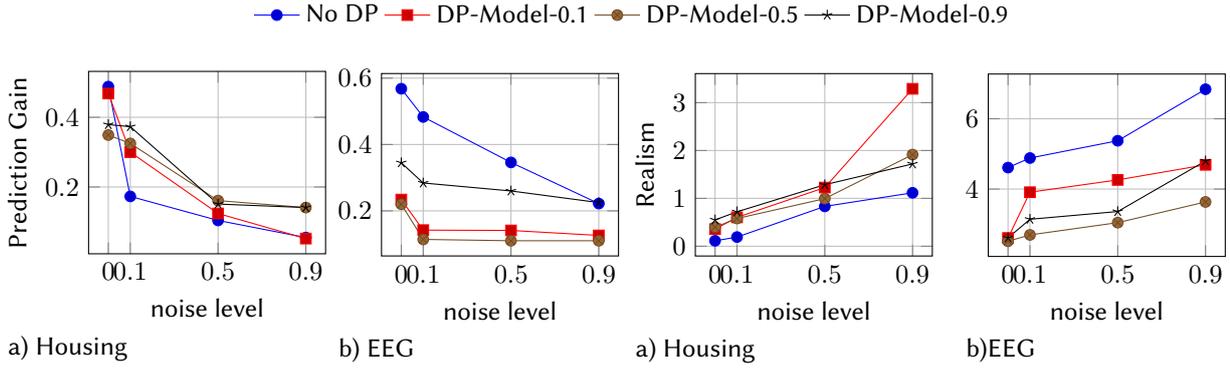
\begin{figure}[ht]
    % First Plot: Housing
    \hspace{-20pt}
    \begin{minipage}{0.24\textwidth}
        \centering
        \begin{tikzpicture}
      %  \hspace{-25pt}
            \begin{axis}[
                width=4.7cm, height=4cm,
                xlabel={noise level },
                ylabel={Prediction Gain},
                title style={yshift=-1ex},
                legend columns=4,
                legend style={
                at={(2.3,1.2)},         % Slightly above the plot
                anchor=south,
                font=\small,
                draw=none,
                fill=none,
                inner sep=1pt
              },
                grid=major,
                label style={font=\small},
                tick label style={font=\small},
                xtick={0,0.1,0.5,0.9},
            ]
            \addplot coordinates { (0,0.488) (0.1,0.173) (0.5,0.104) (0.9,0.055) };
            \addlegendentry{No DP}

            \addplot coordinates { (0,0.468) (0.1,0.3) (0.5,0.124) (0.9,0.052) };
            \addlegendentry{DP-Model-0.1}

            \addplot coordinates { (0,0.349) (0.1,0.325) (0.5,0.161) (0.9,0.141) };
            \addlegendentry{DP-Model-0.5}

            \addplot coordinates { (0,0.379) (0.1,0.373) (0.5,0.151) (0.9,0.141) };
            \addlegendentry{DP-Model-0.9}
            \end{axis}
            \node[anchor=north west] at (current bounding box.south west) 
            {\parbox{\linewidth}{\raggedright \small{a) Housing}}};
        \end{tikzpicture}
    \end{minipage}
    % Second Plot: EEG
    \hspace{10pt}
    \begin{minipage}{0.24\textwidth}
        \centering
        \vspace{25pt}
        \begin{tikzpicture}
            \begin{axis}[
                width=4.7cm, height=4cm,
                xlabel={noise level},
                title style={yshift=-1ex},
                grid=major,
                label style={font=\small},
                tick label style={font=\small},
                xtick={0,0.1,0.5,0.9},
            ]
            \addplot coordinates { (0,0.568) (0.1,0.483) (0.5,0.346) (0.9,0.222) };
            %\addlegendentry{Model Noise = 0}

            \addplot coordinates { (0,0.234) (0.1,0.142) (0.5,0.141) (0.9,0.126) };
            %\addlegendentry{Model Noise = 0.1}

            \addplot coordinates { (0,0.22) (0.1,0.114) (0.5,0.110) (0.9,0.11) };
            %\addlegendentry{Model Noise = 0.5}

            \addplot coordinates { (0,0.345) (0.1,0.284) (0.5,0.26) (0.9,0.225) };
            %\addlegendentry{Model Noise = 0.9}
            \end{axis}
            \node[anchor=north west] at (current bounding box.south west) 
            {\parbox{\linewidth}{\raggedright \small{b) EEG}}};
        \end{tikzpicture}
    \end{minipage}
    % First Plot: Housing - Realism
    \begin{minipage}{0.24\textwidth}
    \vspace{30pt}
        \begin{tikzpicture}
            \begin{axis}[
                width=4.7cm, height=4cm,
                xlabel={noise level },
                ylabel={Realism},
                title style={yshift=-1ex},
                legend columns=4,
                legend style={
                at={(1.2,1.2)},         % Slightly above the plot
                anchor=south,
                font=\small,
                draw=none,
                fill=none,
                inner sep=1pt
              },
                grid=major,
                label style={font=\small},
                tick label style={font=\small},
                xtick={0,0.1,0.5,0.9},
            ]
            \addplot coordinates { (0,0.113) (0.1,0.191) (0.5,0.832) (0.9,1.116) };
            %\addlegendentry{No DP}

            \addplot coordinates { (0,0.356) (0.1,0.603) (0.5,1.225) (0.9,3.289) };
            %\addlegendentry{DP-Model-0.1}

            \addplot coordinates { (0,0.397) (0.1,0.579) (0.5,0.994) (0.9,1.914) };
            %\addlegendentry{DP-Model-0.5}

            \addplot coordinates { (0,0.55) (0.1,0.719) (0.5,1.285) (0.9,1.719) };
            %\addlegendentry{DP-Model-0.9}
            \end{axis}
            \node[anchor=north west] at (current bounding box.south west) 
            {\parbox{\linewidth}{\raggedright \small{a) Housing}}};
        \end{tikzpicture}
    \end{minipage}
    \hspace{10pt}
    % Second Plot: EEG - Realism
    \begin{minipage}{0.24\textwidth}
    \vspace{30pt}
        \centering
        %\vspace{33pt}
        \begin{tikzpicture}
            \begin{axis}[
                width=4.7cm, height=4cm,
                xlabel={noise level},
                title style={yshift=-1ex},
                grid=major,
                label style={font=\small},
                tick label style={font=\small},
                xtick={0,0.1,0.5,0.9},
            ]
            \addplot coordinates { (0,4.611) (0.1,4.883) (0.5,5.375) (0.9,6.841) };
            %\addlegendentry{Model Noise = 0}

            \addplot coordinates { (0,2.607) (0.1,3.91) (0.5,4.26) (0.9,4.69) };
            %\addlegendentry{Model Noise = 0.1}

            \addplot coordinates { (0,2.515) (0.1,2.695) (0.5,3.044) (0.9,3.633) };
            %\addlegendentry{Model Noise = 0.5}

            \addplot coordinates { (0,2.61) (0.1,3.14) (0.5,3.36) (0.9,4.8) };
            %\addlegendentry{Model Noise = 0.9}
            \end{axis}
            \node[anchor=north west] at (current bounding box.south west) 
            {\parbox{\linewidth}{\raggedright \small{b)EEG}}};
        \end{tikzpicture}
    \end{minipage}
\caption{Prediction Gain and Realism achieved by explainers in the Housing and EEG datasets under various DP-Explainer noise levels.}
\label{PredictionGain}
\end{figure}

Figure \ref{PredictionGain} shows the prediction gain achieved by explainer across the Housing and EEG datasets for varying noise level. In the Housing dataset, a clear trend emerges as the DP-Explainer noise increases the prediction gain decreases, which means that employing more noise decreases the CF probability toward the desired class. For example, for No DP, the prediction gain starts at 0.488. However, for DP-Explainer with noise level of 0.9 is applied, it drops dramatically to 0.055. This decline is observed consistently across all model noise levels. Moreover, for DP-Model noise levels (0.5 and 0.9) are introduced, the prediction gain prediction observed is less than that of no DP and DP-model 0.1, regardless of the DP-Explainer noise. Similarly, the EEG dataset follows a comparable pattern. In scenarios without DP applied to the model, the prediction gain is lower and ranges from 0.568 to 0.222 as the DP-Explainer noise is higher. When the model is subjected to DP noise at levels of 0.1, 0.5, and 0.9, the prediction gains are consistently lower.  We now focus on analyzing the impact of incorporating DP on realism. Across both datasets, increasing the DP-Explainer noise consistently results in higher realism scores, indicating less realistic CFs and degradation in CF quality. In the Housing dataset, even without any DP-model noise, the realism score ranges from 0.113 to 1.116 at a DP-explainer noise of 0.9. This degradation is further amplified when additional DP-Model noise is introduced, e.g. with a DP-Model noise of 0.1, the realism score ranges from 0.356 to 3.289 as DP-explainer noise is higher, and similar patterns are observed for DP-Model-0.5 and 0.9. The EEG dataset exhibits a comparable pattern, although the No DP realism scores are generally higher.

\textbf{Discussion on Performance-Privacy-Explanations Interplay:} Results indicate that introducing DP mechanisms affects model performance, although the extent of this impact varies according to the specific use case and dataset. Similarly, the quality of the generated CF explanations is influenced by the privacy parameters applied. Experiments reveal that even slight amounts of noise, whether introduced at DP-Model or within the DP-Explainer, can alter CF quality. In terms of the effectiveness of DP interventions in the context of MEA. Analysis shows that introducing minimal noise at the model level generally offers resistance to MEA. In contrast, higher noise levels provide a more robust defense, albeit at the cost of reduced model performance. When examining the impact of noise on the CFs, we observe that small increments in noise can slightly reduce the success rate of MEA, but further increases yield a more pronounced protective effect. Notably, when both the model and the explainer are simultaneously subjected to DP, a synergistic improvement in resistance to MEA is observed.
\vspace{-10pt}
\section{Conclusion}\label{conclusion}
In this work, we investigate the impact of differential privacy (DP) in mitigating model extraction attacks (MEAs) that leverage counterfactual explanations (CFs) within Machine Learning as a Service (MLaaS) environments. We evaluate employing DP implemented at the ML model level via DP-Stochastic Gradient Descent and at the explanation level, and at both simultaneously, to investigate their respective impacts on MEA resilience. Our analysis, conducted across two datasets, demonstrates and quantifies a fundamental trade-off between privacy protection and utility. The introduction of DP noise presents a clear trade-off, as it effectively hinders an adversary's ability to reconstruct the target model, yet it simultaneously compromises both model performance and the quality of generated CF. Further research will include testing other DP-based methods to generate CFs, MEA methods and more datasets.
\vspace{-10pt}
\bibliography{sample-ceur}
\end{document}